\pdfoutput=1

\documentclass[cernpreprint,texlive=2011,texmf,txfonts,UKenglish]{atlasdoc}

\usepackage[subfigure=false,biblatex=false]{atlaspackage} 
\usepackage{atlasphysics} 
\usepackage{lipsum}
\usepackage{graphicx}
\usepackage{shortcuts} 
\usepackage{slashbox}
\usepackage{cite}%

\newcommand{\papertitle}{Determination of the top-quark pole mass using $t \bar{t}+1$-jet events collected with the ATLAS experiment in 7 TeV $pp$ collisions}
\AtlasTitle{\papertitle}
\AtlasJournal{JHEP}

\PreprintIdNumber{CERN-PH-EP-2015-100}

\AtlasAbstract{

The normalized differential cross section for top-quark pair production in association with at least one jet is studied as a function of the inverse of the invariant mass of the $t \bar{t}+1$-jet  system.  
This distribution can be used for a precise determination of the top-quark mass since gluon radiation depends on the mass of the quarks. The experimental analysis is based on proton--proton collision data collected by the ATLAS detector at the LHC with a centre-of-mass energy of 7 TeV corresponding to an integrated luminosity of 4.6~fb$^{-1}$. The selected events were identified using the lepton+jets top-quark-pair decay channel, where lepton refers to either an electron or a muon. The observed distribution is compared to a theoretical prediction at next-to-leading-order accuracy in quantum chromodynamics using the pole-mass scheme. With this method, the measured value of the top-quark pole mass, $m_t^{\rm pole}$, is:

\[
m_t^{\rm pole}=173.7 \pm 1.5 ~({\rm stat.}) \pm 1.4 ~({\rm syst.})~ ^{+ 1.0}_{-0.5} ~({\rm theory})~{\rm GeV}.
\]

This result represents the most precise measurement of the top-quark pole mass to date.

}

\begin{document}

\maketitle
\nolinenumbers

\section{Introduction}
\label{sec:intro}

In the Standard Model (SM) of particle physics the couplings of the top quark to other particles are fixed through the gauge structure. 
The only free parameters in the top-quark sector of the SM are the elements of the Cabibbo--Kobayashi--Maskawa mixing matrix and the top-quark mass. Due to its high value compared to the other quark masses, accurate knowledge of the top-quark mass is particularly relevant because it is related to the Higgs-boson and $W$-boson masses through radiative and loop corrections. The precise determination of these quantities allows a stringent test of whether the model is consistent \cite{Baak:2014ora,Moch:2014tta}. In addition, the  precise knowledge of the top-quark mass is a crucial ingredient in recent evaluations of the stability of the electroweak vacuum~\cite{Degrassi:2012ry,Alekhin:2012py,Buttazzo:2013uya}.     
  
The top-quark mass was determined directly at the Tevatron and at the Large Hadron Collider (LHC). A combination of a subset of these measurements yields a value of $\mt=173.34\pm0.76$ \GeV~\cite{ATLAS-CONF-2014-008}.  In these measurements, the top-quark mass is inferred from a kinematic reconstruction of the invariant mass of its decay products which is then calibrated to the mass definition used in the Monte Carlo (MC) simulations. These \mt determinations lack a clear interpretation in terms of a well-defined top-quark mass theoretical scheme as employed in quantum chromodynamics (QCD) perturbative calculations,  electroweak fits or any theoretical prediction in general~\cite{Moch:2014tta, ATLAS-CONF-2014-008, ahoang08, GmpLHC:2011bbs, ahoang14}. The values extracted using these methods are usually identified with the top-quark pole mass \mtPole, but present studies estimate differences between the two top-quark mass definitions of $O$(1) GeV~\cite{Moch:2014tta, ATLAS-CONF-2014-008, ahoang08, GmpLHC:2011bbs, ahoang14}.

The top-quark mass can also be measured from the inclusive cross section for top-quark pair (\ttbar) production~\cite{Langenfeld:2009wd}. With this method the top-quark mass scheme is unambiguously defined in the theoretical calculations. However, top-quark mass determinations based on cross-section measurements are less precise, in their current form, than the other techniques based on kinematic reconstruction. This is due to a relatively weak sensitivity of the inclusive top-quark pair production cross section to the top-quark mass, as well as to the large uncertainties on the factorization and renormalization scales and the proton parton distribution function (PDF). To date, the most precise measurement of this type is based on the 7 TeV and 8 TeV data samples collected by the ATLAS experiment during the years 2011 and 2012, which yields $\mtPole=172.9 ^{+2.5}_{-2.6}$~GeV \cite{Aad:2014kva}. The results from the CMS experiment using this technique only include data collected during 2011 \cite{Chatrchyan:2013haa, Chatrchyan:2013haaCor}.

In this paper the method described in \Ref{Alioli:2013mxa} is followed. The top-quark mass is extracted from a measurement of the normalized 
differential cross section $\n3(\mtPole,\rhos)$ for  \ttbar production with at least one additional jet, \ttbaronejet, as a function of the inverse of the invariant mass of the 
\ttbaronejet system, $\rhos \propto 1/\sqrt{\invttj}$. This distribution is sensitive to the top-quark mass because the amount of gluon radiation depends on its value, with large effects in the phase-space region relatively close to the \ttbaronejet production threshold. This method combines the rigorous interpretation of the mass inferred from the inclusive cross section with the advantage of a greater sensitivity.

The measurement is performed using 7 TeV proton--proton $(pp)$ collision data collected by the ATLAS experiment~\cite{Aad:2008zzm}, corresponding to an integrated luminosity of
4.6~\ifb{} with a total uncertainty of $\pm1.8\%$~\cite{Aad:2013ucp}. Events are selected by using the ``lepton+jets'' final state to identify the \ttbar system and at least one 
additional jet. In this channel one $W$ from the top decay produces a lepton (electron or muon) and a neutrino whereas the other $W$ produces a pair of light quarks. Events are thus required to have exactly one lepton, two jets identified as $b$-quark jets, at least three additional  
jets not identified as $b$-quark jets and a significant amount of missing transverse momentum (\met)
due to the neutrino that escapes detection. The differential cross section is
corrected to the parton level (a procedure referred to as unfolding in the following) and normalized. The top-quark pole mass 
is then extracted through a fit to the data using the predicted normalized differential \ttbaronejet cross section from a next-to-leading-order calculation combined with parton showering (NLO+PS) 
\cite{Alioli:2013mxa,Dittmaier:2007wz,Dittmaier:2008uj, Alioli:2011as}.

\section{Definition of the observable}
\label{sec:n3theory}

The method to extract the top-quark pole mass followed here, proposed in \Ref{Alioli:2013mxa}, exploits the fact that the top-quark mass dependence of the \ttbaronejet cross section, \sigmattj, is enhanced in the phase-space region relatively close to the \ttbaronejet production threshold. This method uses the predictions for \ttbaronejet production at hadron colliders at NLO accuracy reported in \Refs{Dittmaier:2007wz,Dittmaier:2008uj}. A well-defined top-quark pole mass can be extracted by comparing these calculations with the measurement of the normalized \ttbaronejet cross section in $pp$ collisions as a function of the inverse of the invariant mass $\sqrt{\invttj}$ of the \ttbaronejet system:

\begin{equation}
\label{eq:n3Definition}
\n3(\mtPole,\rhos)= 
\frac{1}{\sigmattj} 
\frac{\textnormal{d}\sigmattj}{\textnormal{d}\rhos}(\mtPole,\rhos),
\end{equation}
where $\rhos$ is defined as
\begin{equation}
  \rhos=\frac{2 m_0} {\sqrt{\invttj}},
\end{equation}

and $m_0$ is an arbitrary constant of the order of the top-quark mass. Here and in the following, $m_0 =$ 170~\gev{} is used. The anti-\kt jet reconstruction algorithm~\cite{Cacciari:2008gp,Cacciari:2011ma} is employed to reconstruct the jets. The extra jet, beyond those which originated from the \ttbar decay, is the leading jet with a transverse momentum $\pt >$ 50~\GeV{} and a pseudorapidity $|\eta|<$ 2.5.~\footnote{
ATLAS uses a right-handed coordinate system with its origin at the nominal interaction point (IP) in the centre of the
detector and the $z$-axis along the beam pipe. The $x$-axis points from the IP to the centre of the LHC ring, and the
$y$-axis points upward. Cylindrical coordinates $(r,\phi)$  are used in the transverse plane,
$\phi$ being the azimuthal angle around the beam pipe. The pseudorapidity is defined in terms of the polar angle $\theta$ 
as $\eta=- \ln \left[ \tan{(\theta/2)}\right]$. Transverse momentum and energy are defined as
$\pt=p\sin{\theta}$ and $E_{\mathrm{T}}=E\sin{\theta}$, respectively.} The observable \n3 defined in this way is infra-red safe as demonstrated in the study of \Ref{Alioli:2013mxa}.

In this analysis, the measured normalized and differential cross section, unfolded to parton level, is compared to the theoretical calculations at NLO accuracy, after adding the parton shower evolution (NLO+PS). Including the parton shower is expected to give a better description of the final-state phase space than the NLO calculation alone and is implemented in the publicly available MC generator developed in \Ref{Alioli:2011as}. This generator uses \powheg{} (\powheg-ttJ)~\cite{Frixione:2007vw,Alioli:2010xd,Alioli:2011as} matched with the \pythia{} v8 \cite{SjostranPy8} parton shower. Using a fixed order NLO calculation to fit the data gives a similar \n3-distribution but leads to an estimated top quark pole mass about 0.3 GeV lower than using a NLO+PS calculation.

This difference is well below the present theoretical uncertainty of the calculation. Differences due to the use of \pythia{} v8 or \pythia{} v6~\cite{Sjostrand:2006za} are below this value of 0.3 \GeV{}.

In the NLO calculation, it is assumed that the top quarks are stable. Possible effects due to radiation from top-quark decay products and virtual corrections to the decay are small compared to the overall theoretical uncertainty. Quantum chromodynamics corrections to the decay do not affect the mass renormalization of the top quark at the same order of accuracy as considered in the calculation because the renormalization is purely determined from the QCD self-energy corrections of the top-quark propagator, which is included in the calculation. Furthermore, recent calculations in \Refs{denner12teos,denner12ttos} include NLO QCD corrections to the total and differential \ttbar cross section assuming the top quarks to be off-shell. In this framework, the results in the on-shell approximation are reliable and off-shell effects are at the sub-percent level. In the following, it is assumed that similar results hold for the quantity \n3. In fact, as \n3 is expressed as a normalized cross section, even smaller effects can be expected due to possible cancellations.

\section{The ATLAS experiment}
\label{sec:atlas}

The ATLAS experiment~\cite{Aad:2008zzm} is a multipurpose detector with a forward-backward symmetric cylindrical geometry. The inner tracking detector (ID) consists of a silicon pixel detector,
a silicon microstrip detector (SCT), and a straw-tube transition
radiation tracker (TRT). The ID is surrounded by a thin superconducting
solenoid which provides a 2 T magnetic field and by a high-granularity
liquid-argon (LAr) sampling electromagnetic calorimeter.
The electromagnetic calorimeter is divided into a central barrel
($|\eta| <$ 1.475) and end-cap regions at each end of the barrel (1.375 $< |\eta| <$ 2.5 for the outer wheel and
2.5 $< |\eta| <$ 3.2 for the inner wheel). 
A steel/scintillator-tile calorimeter completes the measurement of hadronic showers in the central pseudorapidity range ($|\eta| <$ 1.7), while
a LAr hadronic end-cap calorimeter provides coverage over 1.5 $<
|\eta| <$ 3.2. The forward regions (3.2 $< |\eta| <$ 4.9) are instrumented
with LAr calorimeters for electromagnetic and hadronic measurements.
The muon spectrometer (MS) surrounds the calorimeters
and consists of three large air-core superconducting toroid systems providing bending powers
of 3 Tm in the barrel and 6 Tm in the end-caps, a system of precision
tracking chambers ($|\eta| <$ 2.7), and fast detectors for triggering ($|\eta| <$ 2.4). The
combination of all these sub-detectors provides charged-particle measurements
together with efficient and precise identification of leptons and photons in the pseudorapidity range of $|\eta| <$ 2.5. Jets and \met are reconstructed using energy deposits over the full 
coverage of the calorimeters, $|\eta| < 4.9$. The reconstructed muon momenta are also included in the evaluation of \met. Evaluation of the luminosity scale
is performed using several luminosity-sensitive detectors. The ATLAS experiment has a three-level trigger 
system~\cite{Aad:2008zzm}. The first-level trigger is hardware-based and uses a subset of the detector information to reduce the 
event rate to at most 75 kHz. The second and third levels are software-based and together reduce the 
event rate to about 300 Hz.

\section{Data sample and Monte Carlo simulation}
\label{sec:sample_data}

The data considered in this analysis correspond to an integrated luminosity of 
4.6 \ifb{} of proton--proton collisions at a centre-of-mass energy of 7~\TeV.
They were recorded in 2011
during periods with stable beam conditions and with all relevant subdetector systems 
operational. The events were selected by single-lepton triggers 
that require a minimum transverse momentum of 18 \gev{} for muons and a 
minimum of 20 to 22~\gev{} for electrons, depending on the data-taking conditions.  


In this analysis, several MC samples are used for the modelling of \ttbar pair production and the main background processes. For the simulation 
of the \ttbar{} signal and the unfolding, the \powheg{} code (\powheg-hvq, {\it patch4}~\cite{Frixione:2007nw}) is used to calculate the QCD matrix 
element at NLO with the CT10~\cite{Lai:2010vv} PDF set. The parton shower and the underlying event are added 
using the \pythia{} v6.4~\cite{Sjostrand:2006za} generator with the Perugia 2011C parameter set (tune)~\cite{Skands:2010ak}. Several MC sets of events  were generated using 
different top-quark masses. The nominal MC sample, which is used in the present study to compare the MC predictions with data, is the largest and
is produced assuming a top-quark mass of $\mt= 172.5$ \GeV. The corresponding cross section of the nominal \ttbar sample 
is $177^{+10}_{-11}$ (theo.) pb as predicted by the calculations in \Refs{Cacciari:2011hy,Baernreuther:2012ws,Czakon:2012pz,Czakon:2013goa,Czakon:2011xx}, 
which include the next-to-next-to-leading order (NNLO) and the resummation of next-to-next-to-leading logarithmic (NNLL) soft gluon terms with top++2.0~\cite{Czakon:2013goa}. In addition 
to this nominal sample, five other samples are employed with different assumptions about the input top-quark mass in the range from $167.5~\GeV$ to $180~\GeV$ in steps of $2.5$ \GeV.

Background processes to the \ttbaronejet final state under study are simulated with various MC generators. Single top quark production in the $s$-, $Wt$- and $t$-channels is simulated using
\powheg{} matched with \pythia{} v6.4. The Perugia 2011C tune is used. 
The production of $W/Z$ bosons in association with jets ($W$+jets or $Z$+jets) 
is simulated using the ALPGEN generator (v2.13) \cite{alpgen}  with the 
leading-order (LO) CTEQ6L1~\cite{cteq6} PDF set. These calculations are interfaced with 
\herwig{}~\cite{Corcella:2000bw} for the parton shower and \jimmy{} v4.31~\cite{Butterworth:1996zw}  for the underlying-event modelling. $W$+jets events containing heavy-flavour 
quarks are generated separately using leading-order matrix elements with massive $b$- and $c$-quarks. Possible double-counting due to heavy quarks 
produced by the parton shower is considered and removed. The total number of $W$+jets events is normalized by exploiting the lepton charge asymmetry observed in data, 
following the method in \Ref{ATLAS:2012an}. Diboson events $(WW$, $ZZ$, $WZ)$ are generated using \herwig{} with the MRSTMCal~\cite{Sherstnev:2007nd} PDF. The background from 
misidentified and non-prompt leptons is estimated using a data-driven matrix method described in \Ref{raey}.

Multiple soft $pp$ interactions generated with \pythia{} v.6.425 using the AMBT2B tune~\cite{ATL-PHYS-PUB-2011-009} are added to all simulated events in order to account for the effect of multiple $pp$ interactions in the same and neighbouring bunch crossings (pile-up).

The response of the ATLAS detector is simulated using a detailed description of the detector geometry~\cite{Aad:2010ah} in GEANT4~\cite{geant4}. Simulated events are reconstructed using the same software as used for the data. 

\section{Object definition and basic selection}
\label{sec:objects}

The analysis applies several requirements to the events and makes use of reconstructed electrons, muons, jets and \met. Electron candidates are reconstructed from energy deposits in the electromagnetic 
calorimeter using criteria based on the shower shape, and they must be matched to a charged-particle track in the ID~\cite{Aad:2014fxa}. Electrons must have a transverse momentum of \pt > 25 \gev{} and $|\eta|$ < 2.47.
Events with electrons falling in the calorimeter barrel/end-cap transition region, corresponding to 1.37 $< |\eta| <$ 1.52, are rejected. Muon candidates are identified by combining track candidates in the MS and the ID~\cite{Aad:2014zya}. All muons are required to have a transverse momentum \pt > 25 \gev{} and $|\eta|$ < 2.5. All muons must additionally satisfy a series of selection criteria on the number of hits per track in the various tracking sub-detectors~\cite{Aad:2014zyy}. Finally, electrons and muons have to match corresponding objects that have fired the trigger.

Isolation criteria are applied to electron and muon candidates to reduce the background from hadrons mimicking lepton signatures and backgrounds from heavy-flavour decays inside jets~\cite{Aad:2011yb}.

Energy deposits in the calorimeters are combined into three-dimensional clusters~\cite{Lampl:1099735}. From these clusters, jets are reconstructed using the anti-\kt jet algorithm with a radius parameter of 0.4. Reconstructed jet energies in simulations are calibrated from stable-particle jets. Residual calibrations, derived by using in situ methods where the jet transverse momentum is compared to that of a reference object (e.g. using $\gamma/Z$+jet events), are then applied to data relative to the simulations~\cite{Aad:2014bia}.

Reconstructed jets must have $\pt> 25$  \gev{} and $|\eta|<2.5$. To suppress the contribution from low-\pt jets originating from pile-up interactions, a jet vertex fraction requirement is applied: at least $75\%$ of the summed scalar \pt of tracks associated with the jet must be due to tracks originating at the event primary vertex. This primary vertex is defined as the vertex with the highest $\sum_{{\rm trk}} (\pt^{\rm trk})^2$ among all candidates with at least five associated tracks (${\rm trk}$) with $\pt^{\rm trk}>0.4~\GeV$~\cite{ATLAS-CONF-2010-069}. Jets containing $b$-hadrons   are identified using a neural network exploiting the long lifetime of $b$-hadrons at a working point resulting in a tagging efficiency of $70\%$ in simulated \ttbar events~\cite{ATLAS-CONF-2012-040, ATLAS-CONF-2012-043, ATLAS-CONF-2012-097}.

The transverse momentum of neutrinos escaping the detector is assumed to be identical to \met, which is reconstructed as the magnitude of the momentum imbalance in the transverse plane as obtained from the negative vector sum of the momenta of all energy deposits. It is reconstructed from topological clusters, calibrated for electromagnetic objects and corrected according to the energy scale of the identified objects. Muons contributions are also included by using their momentum measured in the inner detector and the muon spectometer~\cite{Aad:2012re}.

Events must not contain jets with $\pt$ greater than 
20~\gev{} arising from out-of-time energy deposits or from
energy deposits with a hardware or calibration problem.
Exactly one isolated electron or muon, and at least
five jets are required with  exactly two of the jets tagged as $b$-quark jets. The magnitude of the missing transverse momentum and the transverse
mass of the system formed by the charged lepton 
and the neutrino,\footnote{The $W$-boson transverse mass is defined as $m_{\mathrm{T}}^{W}=\sqrt{2p_{\mathrm{T},{\ell}}~p_{\mathrm{T},\nu}[1-\cos{(\phi_{\ell}-\phi_{\nu})}]}$ where ${\ell}$ and $\nu$ refer to the selected lepton and the neutrino whose information is associated with the \met vector, respectively.} $m_{\mathrm{T}}^{W}$, must both exceed 30~\gev{}.

\section{Reconstruction of the $\ttbar+1$-jet system}
\label{sec:reco}

After the basic selection, a kinematic reconstruction of the events is performed to identify the $W$-boson
and top-quark candidates. The leptonically-decaying $W$ boson is identified with the 
charged lepton and the neutrino, where the longitudinal momentum is inferred using a constraint on the $W$-boson mass. Candidates for the hadronically-decaying $W$ boson are constructed by considering all possible pairs of jets among those not identified as $b$-jets. Accepted events must fulfil  the following conditions:

\begin{equation}
0.9<\alpha \equiv \frac{m_W^{\mathrm{ref}}}{m_{ij}} <1.25,
\label{eq:alpha}
\end{equation}
\begin{equation}
\Delta \kt(i,j) \equiv min(\pt^i,\pt^j)\cdot\Delta R(i,j) < 90~\gev{}, 
\label{eq:beta}
\end{equation}
where the indices $i$ and $j$ run over all jets not identified as $b$-jets, $m_{ij}$ is the invariant mass of jets $i$ and $j$, and $m_W^{\mathrm{ref}}=80.4~\GeV$~\cite{PDG}. All combinations of the two $b$-jets with the $W$-boson candidates are considered as top-quark candidates. Once both requirements are applied, the permutation that minimizes the invariant mass difference of the hadronically and leptonically decaying top-quark candidates is selected. The application of these two requirements reduces the combinatorial background. 

\begin{table}[h]
\begin{center}
\begin{tabular}{|l|crr|}
\hline
                             & \hfill Events   & \multicolumn{2}{c|}{\hfill Uncertainty}   \\
\hline
Signal $(\ttbar,~ \mt=172.5~\GeV)$                      & \hfill 2050      &$\hfill \pm$ & 320  \\
\hline
$W$+jets                       & \hfill 31        & \hfill $\pm$ & ~16   \\
$Z$+jets                       &\hfill 6         & \hfill $\pm$ &  ~~4    \\
Single top ($\mt=172.5~\GeV$)                   & \hfill 62 & \hfill $\pm$&  ~34   \\
$WW$, $ZZ$, $WZ$                   & \hfill 1         & \hfill $\pm$ & ~~1 \\
Misidentified and non-prompt leptons  &\hfill  22        & \hfill $\pm$ &  ~13   \\
\hline
Total Background             & \hfill 121       & \hfill $\pm$ &  ~40   \\
Total Predicted              & \hfill 2170      & \hfill $\pm$ &  320 \\
Data                         & \hfill 2256      &   &  \\
\hline
\end{tabular}
\end{center}
\caption{
Event yields and their uncertainties after the reconstruction of the \ttbaronejet system. The quoted uncertainty values include the total statistical and systematic uncertainties as described in section~\ref{sec:systematics}.
\label{tab:yields_n3_final}
}
\end{table} 

\begin{figure}[!]
 \begin{center} 
    \begin{tabular}{cc}
      \includegraphics[width=0.45\columnwidth]{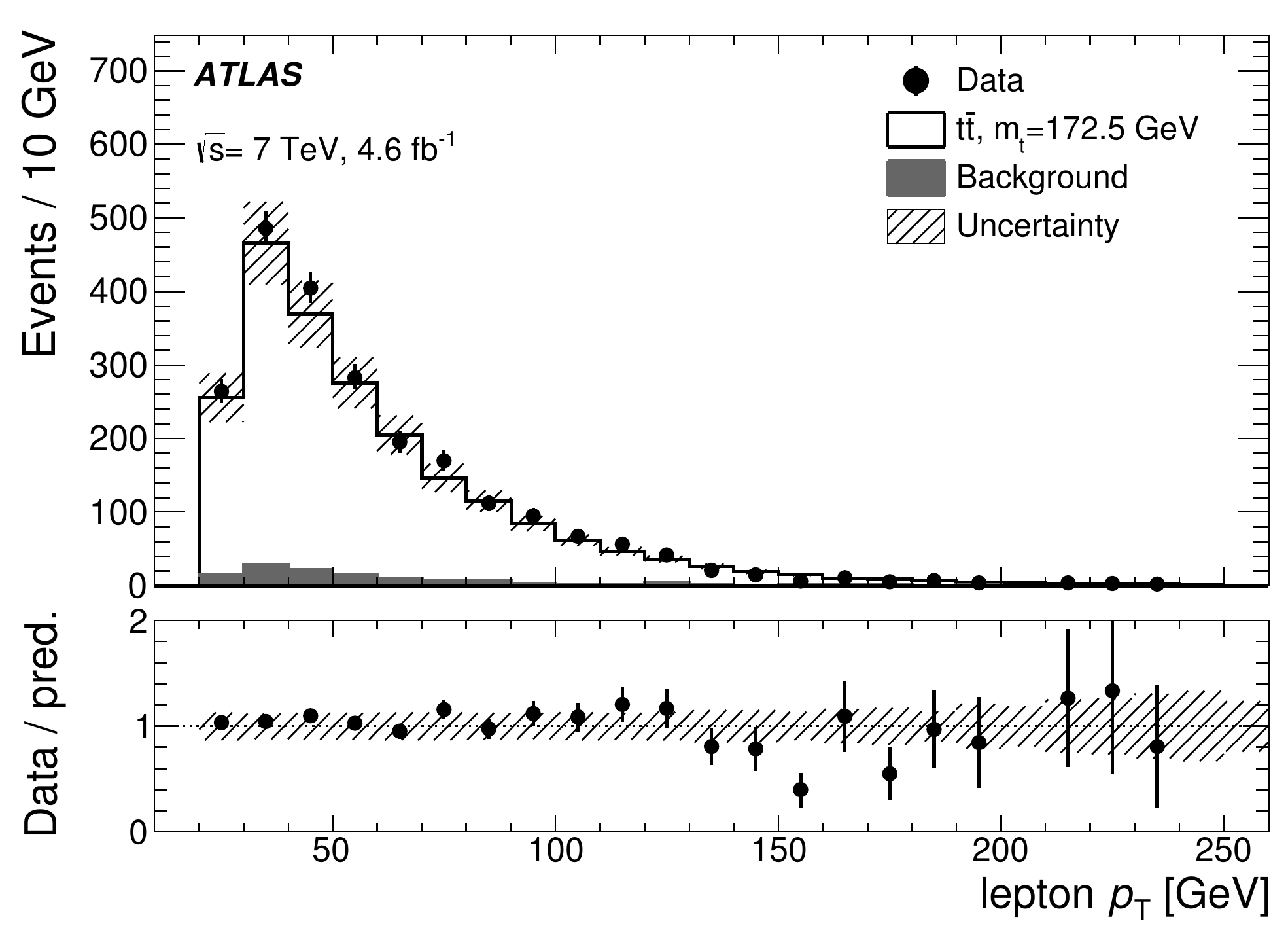}&
      \includegraphics[width=0.45\columnwidth]{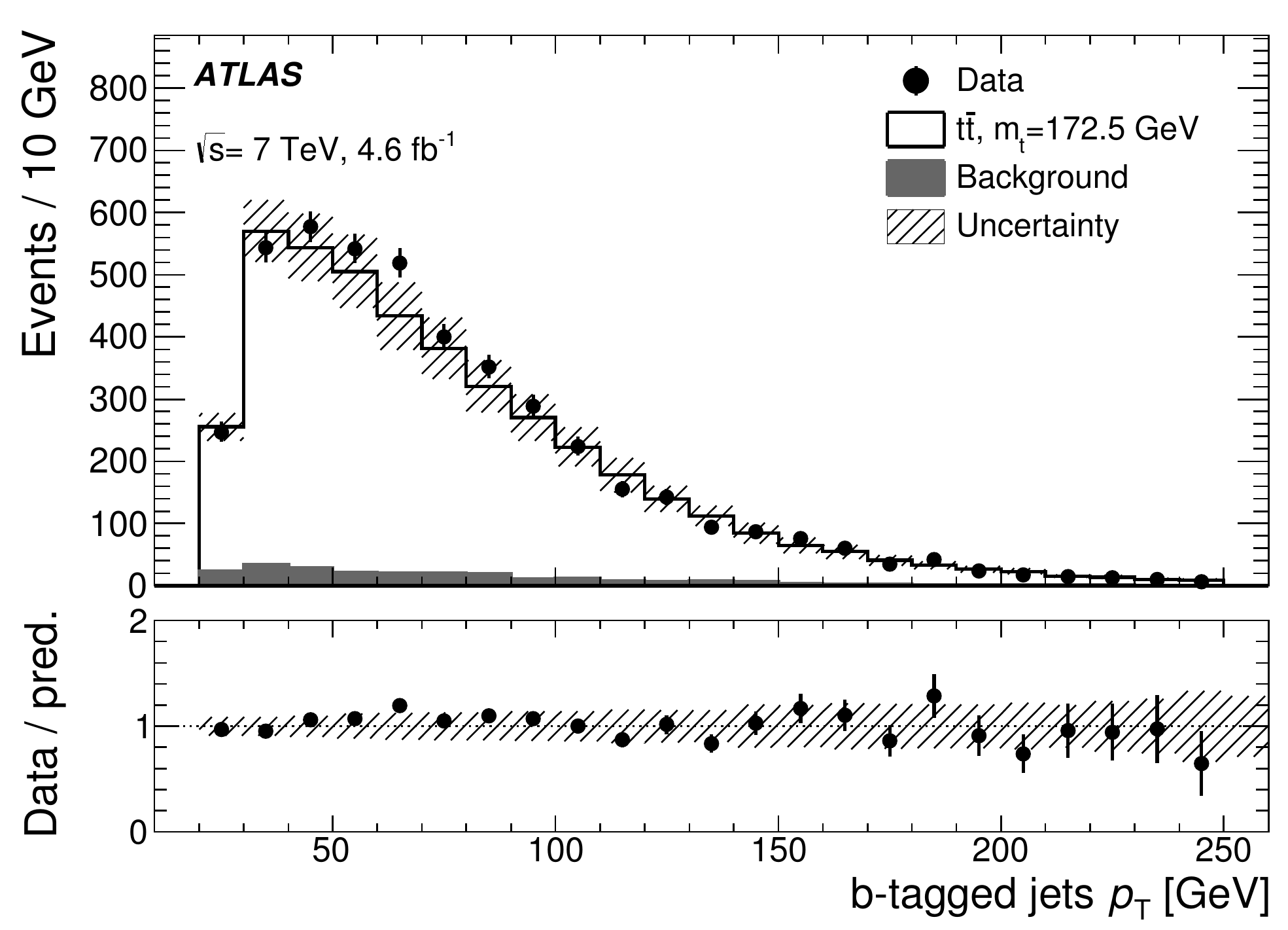}\\
      \includegraphics[width=0.45\columnwidth]{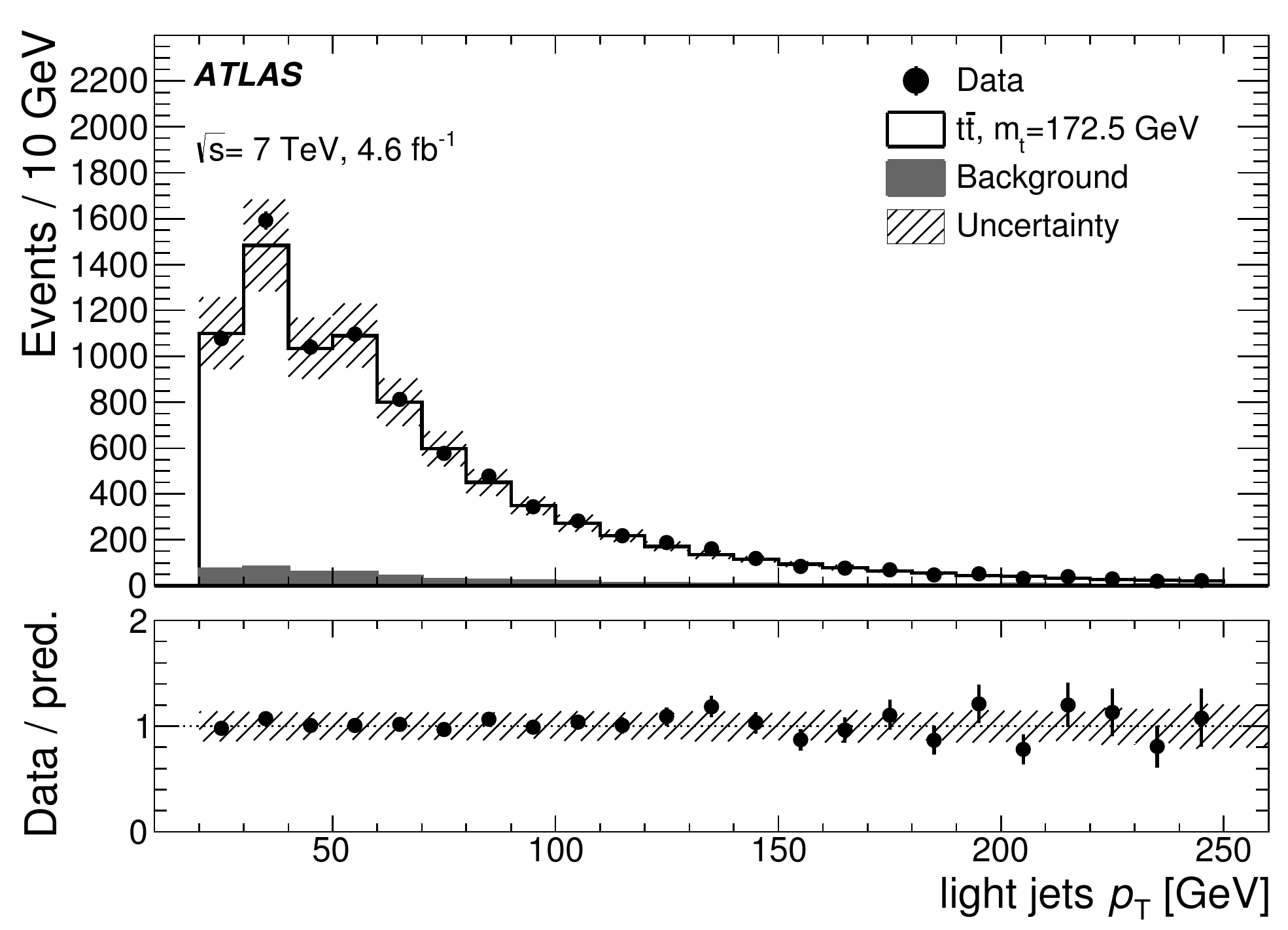}&
      \includegraphics[width=0.45\columnwidth]{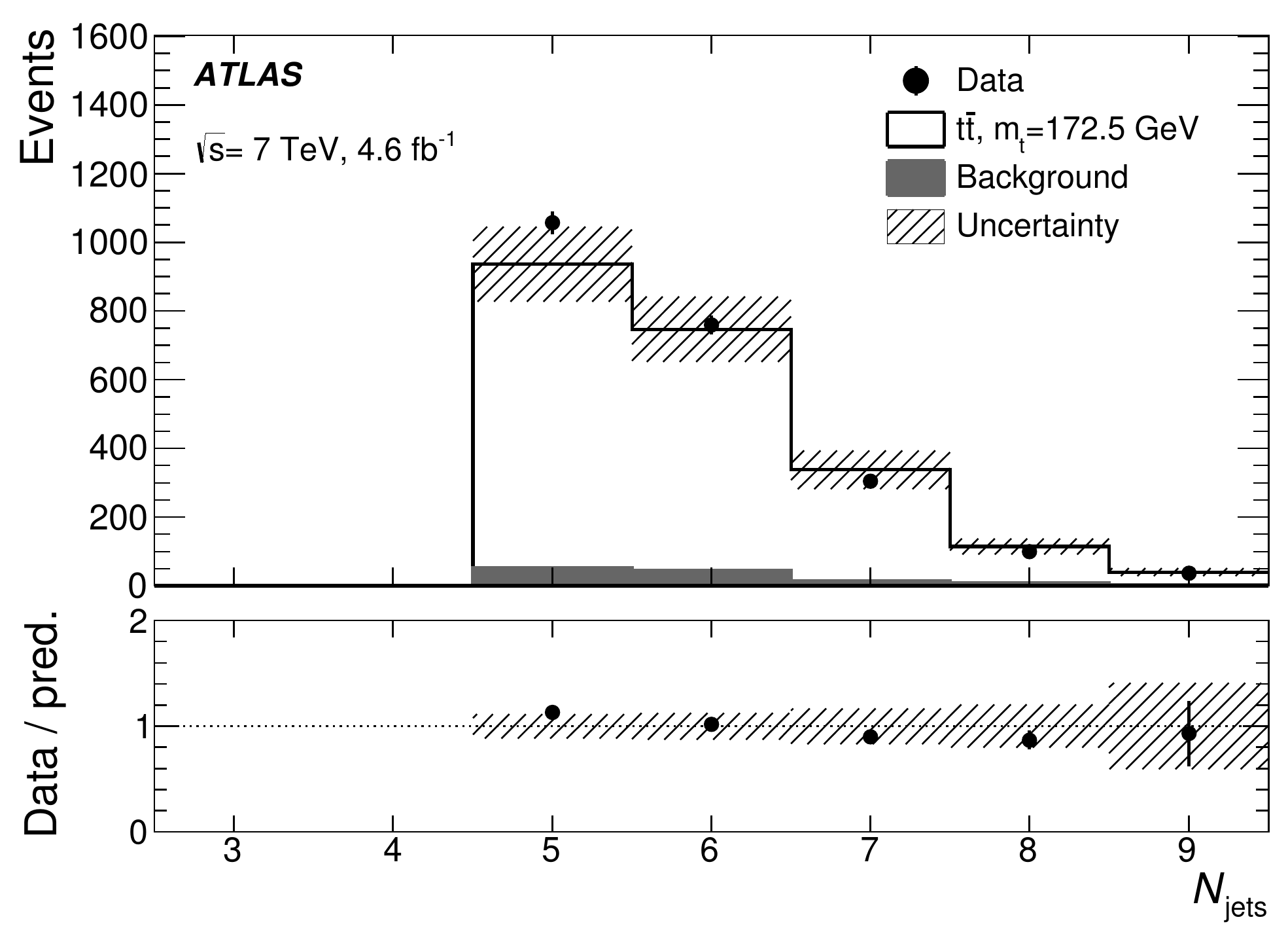}\\
    \end{tabular}
  \end{center}
\caption{The data for various kinematic distributions (transverse momentum, \pt, of the lepton, \pt of all the $b$-tagged jets, \pt of all non-$b$-tagged jets and the total jet multiplicity) are compared to the nominal \ttbar MC sample (\powheg+\pythia) plus backgrounds after the final kinematic reconstruction of the \ttbaronejet events. The total background estimated in table~\ref{tab:yields_n3_final} is shown in dark grey. The uncertainty band includes the total statistical and systematic uncertainties as described in section~\ref{sec:systematics}.
\label{fig:reco}}
\end{figure}
 
\clearpage

\begin{figure}[t]
 \begin{center} 
    \begin{tabular}{cc}
      \includegraphics[width=0.45\columnwidth]{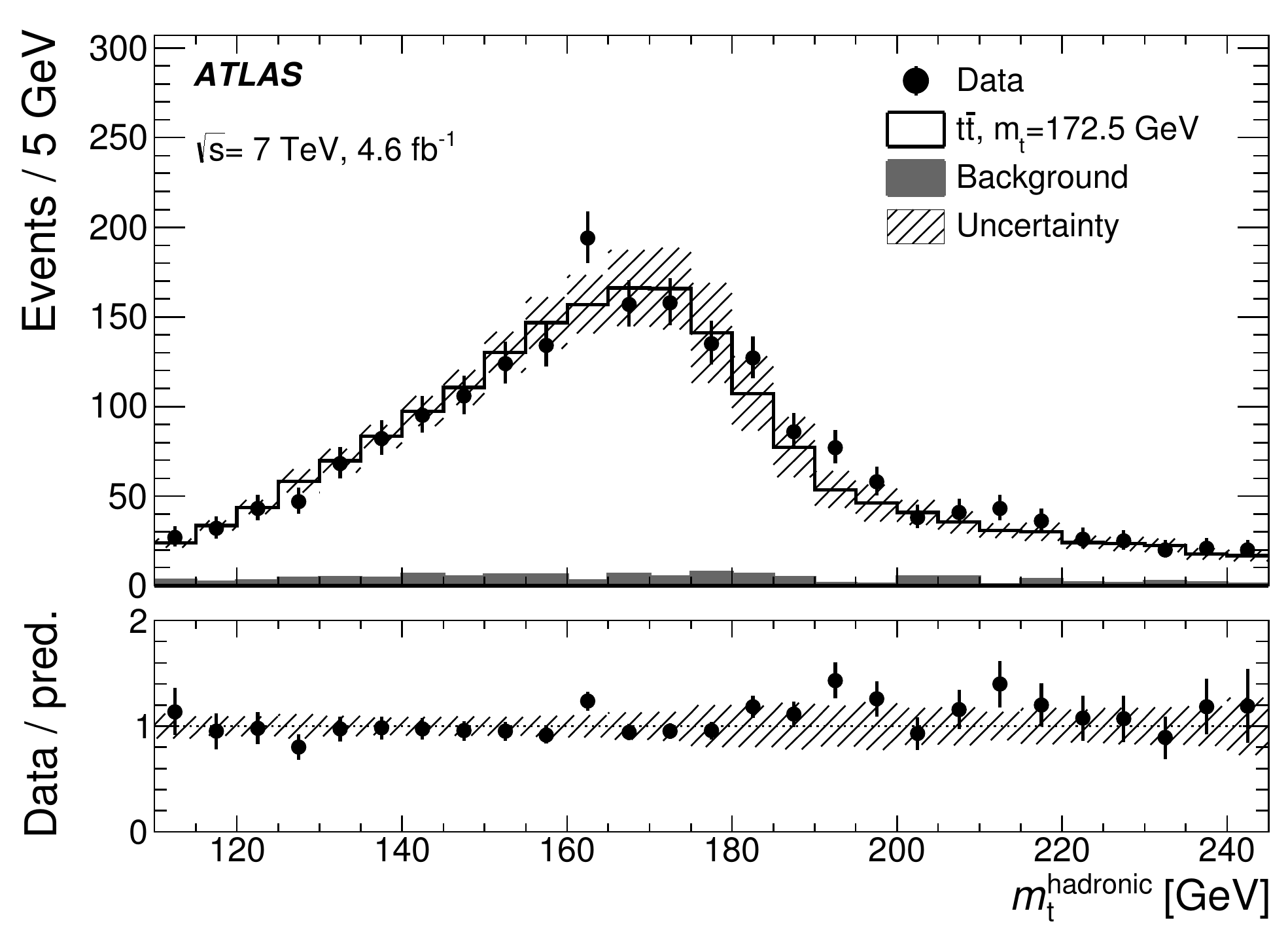} &
      \includegraphics[width=0.45\columnwidth]{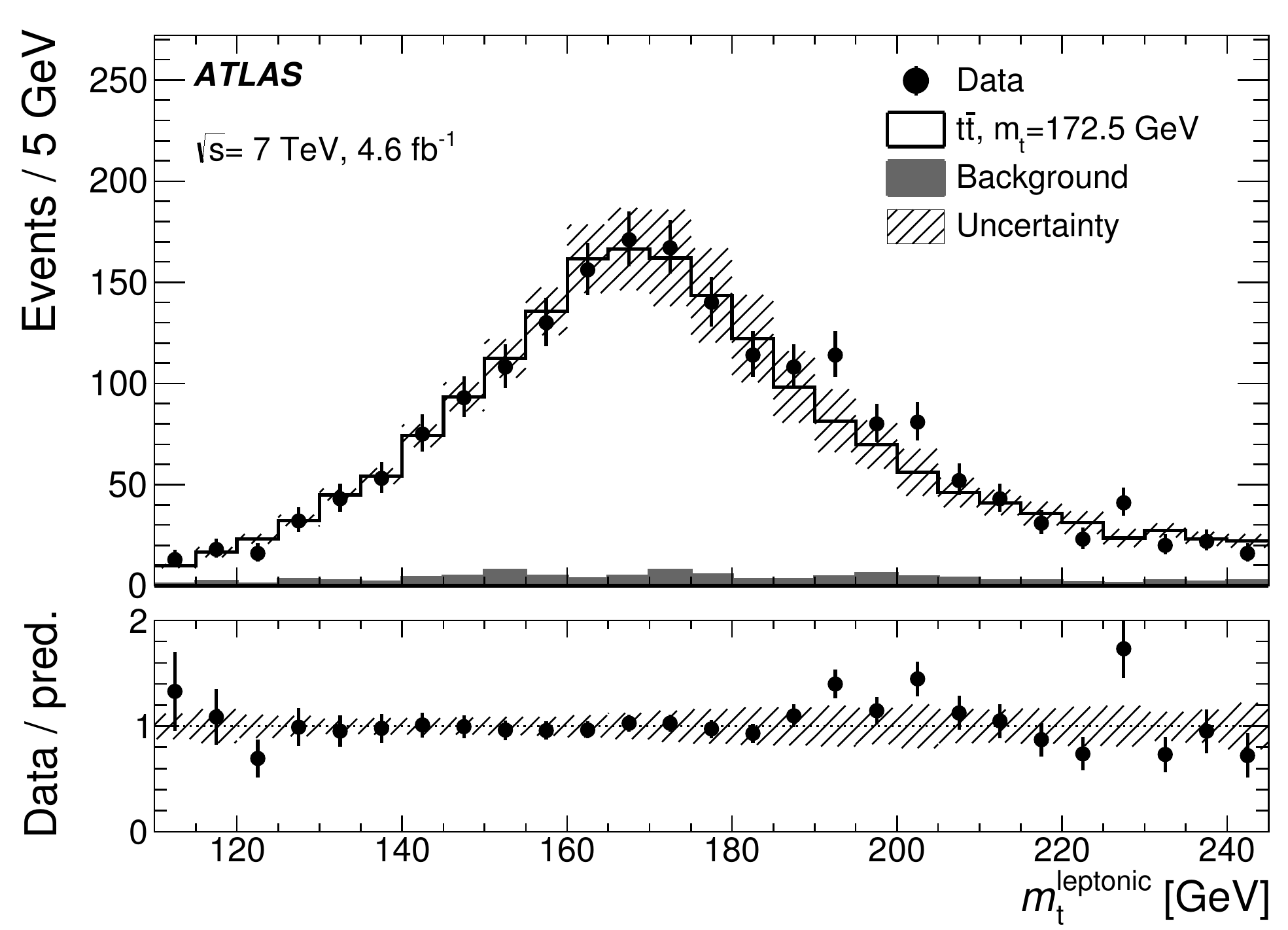} \\
      \includegraphics[width=0.45\columnwidth]{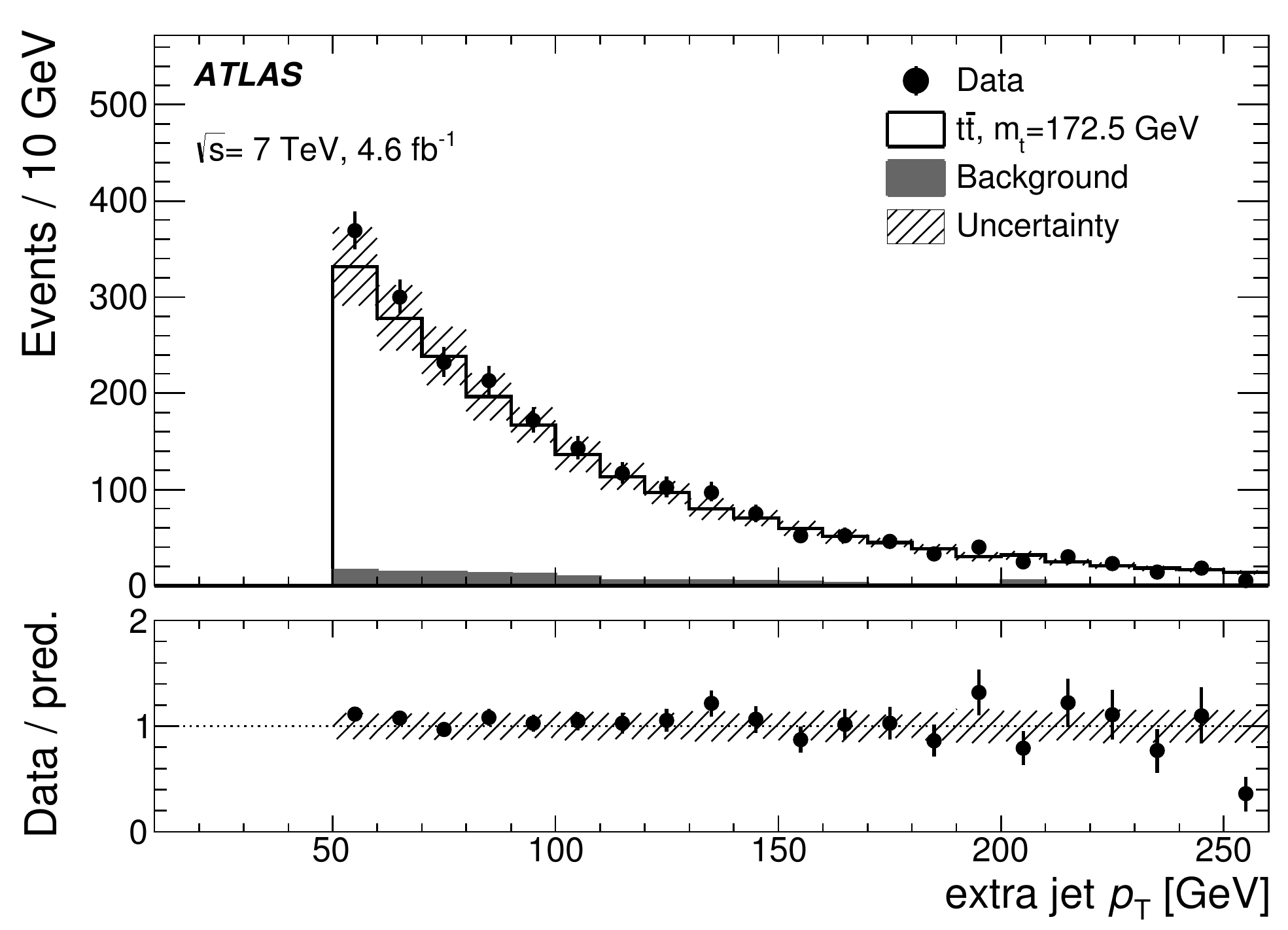} &
      \includegraphics[width=0.45\columnwidth]{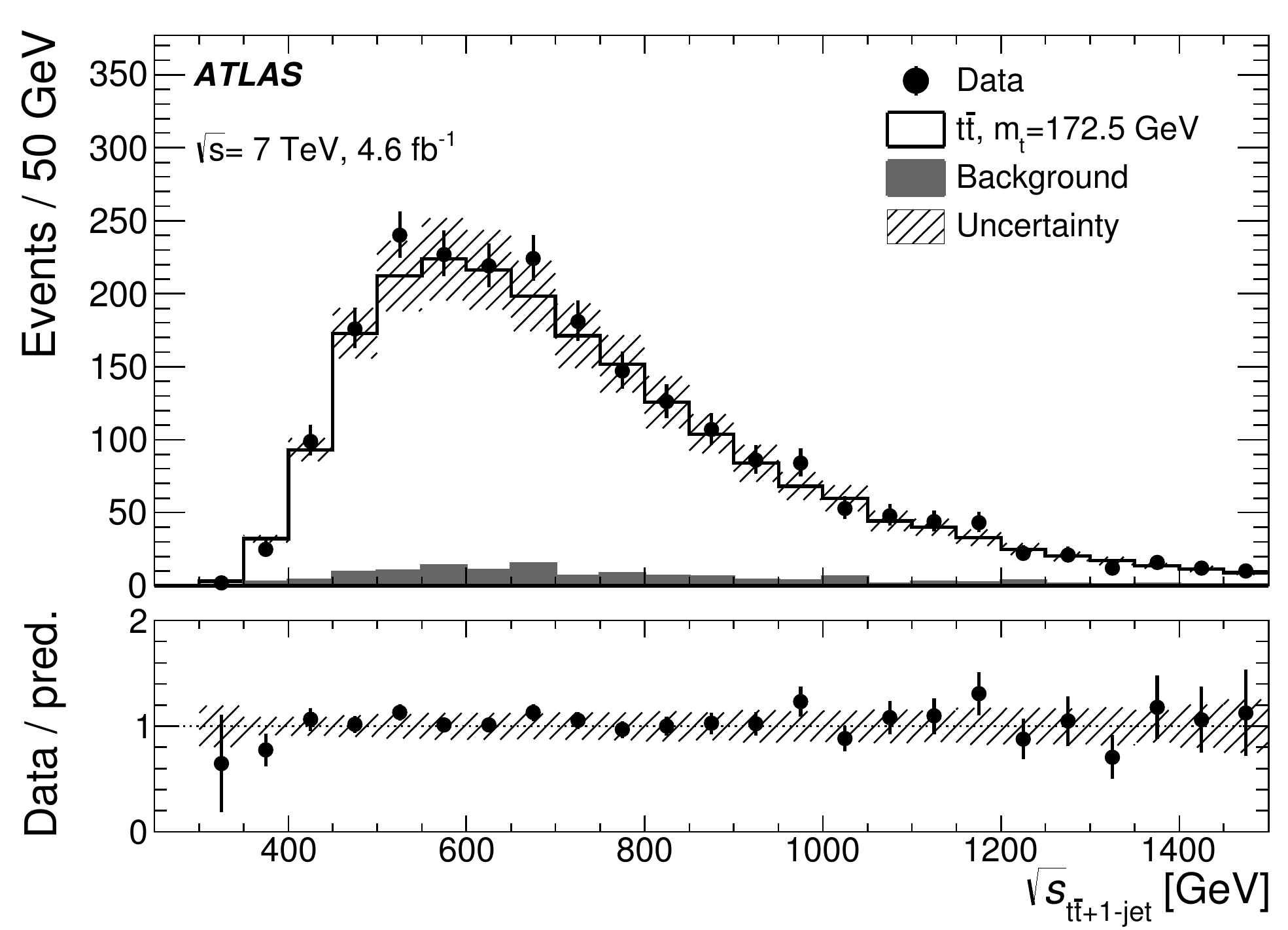}\\
    \end{tabular}
  \end{center}
\caption{The data for various kinematic distributions (the reconstructed mass of the hadronically and leptonically decaying top-quark candidates, the \pt of the additional jet and the invariant mass of the \ttbaronejet system) are compared to the nominal \ttbar MC sample (\powheg+\pythia) plus backgrounds after the final kinematic reconstruction of the \ttbaronejet events. The total background estimated in table~\ref{tab:yields_n3_final} is shown in dark grey. The uncertainty band includes the total statistical and systematic uncertainties as described in section~\ref{sec:systematics}.
\label{fig:reco2}}
\end{figure}

\begin{figure}[t]
\begin{center}
\begin{tabular}{c}
\includegraphics[width=0.8\columnwidth]{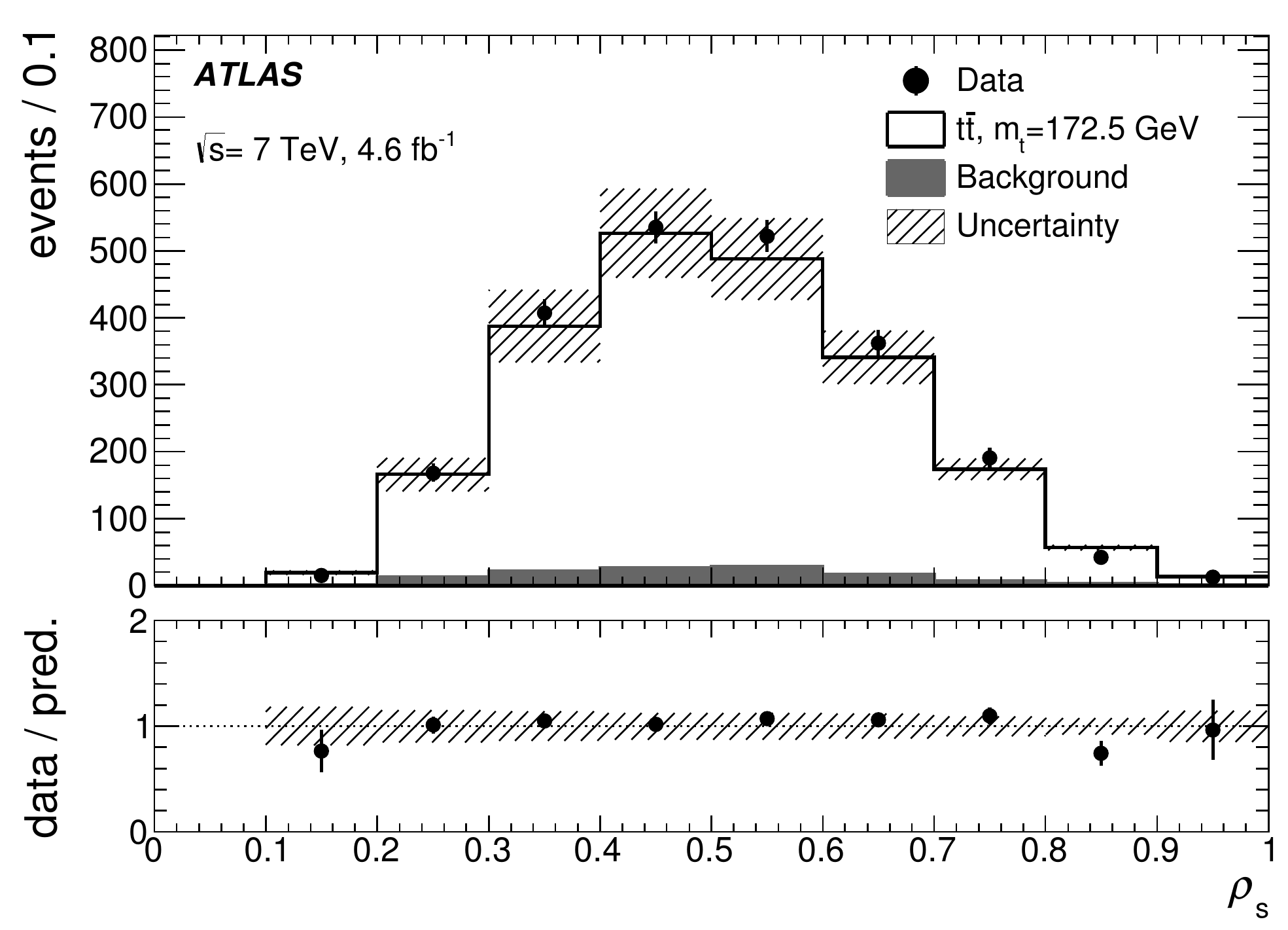}\\
\end{tabular}
\end{center}
\caption{Number of reconstructed events as a function of \rhos ($m_0=170~\GeV$) related to the inverse of the invariant mass of the \ttbaronejet system. The data are compared to the nominal \ttbar MC prediction (\powheg+\pythia) plus backgrounds, which assumes a top-quark mass $\mt = $172.5 \gev{}
after the final kinematic reconstruction of the \ttbaronejet events. The backgrounds and the systematic uncertainties are estimated as in figure~\ref{fig:reco} and~\ref{fig:reco2}.
\label{fig:n3reco}}
\end{figure}

The magnitude of the momentum vectors of the light-quark jets associated with the hadronic $W$ boson 
are scaled using the value of $\alpha$ derived from eq.~\ref{eq:alpha}.  A further requirement on the ratio of the reconstructed 
top-quark kinematic mass for the leptonic and hadronic decays, $\mt^{\mathrm{leptonic}}/\mt^{\mathrm{hadronic}}>0.9$,
is imposed to increase the signal to background ratio. The leading-$\pt$ jet is selected among the remaining jets and is identified as the extra jet completing the \ttbaronejet system. The extra jet must satisfy $\pt >$ 50~\gev{} and $|\eta| <$ 2.5.

The event yields after the final selection are presented in table~\ref{tab:yields_n3_final}. The number
of selected data events is in good agreement with the MC expectation. 

After applying all the selection criteria, data were compared to the expectations. A representative subset of  kinematic and angular distributions is shown in figures \ref{fig:reco} and \ref{fig:reco2}. In all cases, good agreement between data and prediction is observed. In addition to these plots, the number of reconstructed \ttbaronejet events as a function of \rhos is shown in figure \ref{fig:n3reco}.

\section{Top-quark mass determination}
\label{sec:methodology}

The top-quark pole mass is extracted by fitting the parameterized NLO+PS prediction to the measured distribution of the normalized differential 
cross section \n3 defined in eq. \ref{eq:n3Definition} after background subtraction and correction for detector effects and hadronization. This method follows a similar procedure to that employed in \Ref{Aad:2014zka}. The observed number of \ttbaronejet events in figure \ref{fig:n3reco} is used to construct this \n3-distribution at parton level. 

The recorded parton-level information in the nominal ATLAS signal MC sample (see section \ref{sec:sample_data}) did not allow the construction of the extra jet at the parton level, as required by the theoretical calculation. Consequently, a simple direct connection between the \ttbaronejet system at detector level and at parton level could not be made. For that reason, an intermediate state corresponding to the first gluon emission at parton level (\ttbar+g) was introduced to bridge the connection and  an additional MC sample of events was generated. This latter sample transformed the \n3-distribution of the $\ttbar$+g system into the final \n3-distribution of the \ttbaronejet system, which this time was defined as in the theoretical calculation. This sample was validated using ATLAS procedures and did not include the full detector simulation. The final correction procedure contained the following steps:

\vspace*{0.2cm}
\centerline{
\ttbaronejet  (detector level) $\rightarrow$ $\ttbar$+g (parton ~level) $\rightarrow$ \ttbaronejet (parton ~level), 
}

As a first step the migration matrix relating  the distribution of \ttbaronejet events at detector level and the \ttbar+g events at parton level was calculated using the nominal MC simulation. Data were grouped in bins as a function of \rhos with a variable bin size as displayed in table \ref{tab:n3_theory} and in figure \ref{fig:best_fit}. This choice was the result of a compromise between having values of the diagonal terms in the migration matrix above 50\% and optimizing the sensitivity of the \n3-distribution to the top-quark mass according to the study of \Ref{Alioli:2013mxa}. The migration matrix was then inverted using the Singular Value Decomposition (SVD) method described in \Ref{Hocker:1995kb}. This algorithm minimizes the statistical fluctuations inherent in the matrix-inversion process. 

The above correction restricts the kinematical phase-space region to that of the events satisfying the selection criteria. Hence an additional correction is needed to extend this region to the full acceptance considered by the theoretical calculation (acceptance term). Finally  an additional step was implemented to convert the \n3-distribution corresponding to the $\ttbar$+g system into the  \n3-distribution of the \ttbaronejet system defined as in the theoretical calculation. The final unfolding procedure is described as follows:

\begin{equation}
\Rcor (\rhos) \equiv \left[ \left( \mathcal{M}^{-1} \otimes \Rdet(\rhos)\right) \cdot \left( \frac{\Rttg_{\mathrm{acc.}}(\rhos)}{\Rttg (\rhos)}\right)^{-1} \right] \cdot \left(\frac{\Rttj(\rhos)}{\Rttg(\rhos)}\right),
\label{eq:unfolding}
\end{equation}
where $\left( \mathcal{M}^{-1} \otimes \Rdet(\rhos)\right)$ is the term describing the transformation of the \ttbaronejet  distributions from detector level to the parton level at its first gluon emission ($\ttbar$+g). 
The acceptance term is
\begin{equation}
\left(\frac{\n3^{\ttbar+{\rm g}}_{\rm acc.}(\rhos)}{\n3^{\ttbar+{\rm g}}(\rhos)}\right)^{-1},
\label{eq:unfolding_bis}
\end{equation}
and deviates from unity by less than 2\% over the full \rhos range. The factor 
\begin{equation}
\left(\frac{\n3^{\ttbar+1\scriptsize{\textnormal{-\rm{jet}}}}(\rhos)}{\n3^{\ttbar+{\rm g}}(\rhos)}\right),
\label{eq:unfolding_bis2}
\end{equation}
transforms the \n3-distribution which corresponds to the $\ttbar$+g system into the \n3-distribution of the \ttbaronejet system. This correction factor is typically within 10\% of unity.

In more detail, the first part of eq. \ref{eq:unfolding} corresponds to:

\begin{equation}
\left( \mathcal{M}^{-1} \otimes \Rdet(\rhos)\right)_i  \equiv  \frac{1}{\Nttj_{\mathrm{tot}}}  \frac{\sum \limits_{j} {M}^{-1}_{ij} \cdot \Nttj_{j}}{\Delta \rhos^{i}},
\label{eq:unfolding2}
\end{equation}
where $i,j$ refers to the bin numeration defined for the \rhos variable, $\Nttj_{j}$ is the number of \ttbaronejet events reconstructed (background subtracted) in the $j$-th bin, $\Nttj_{\mathrm{tot}}$ is the total number of reconstructed \ttbaronejet events (background subtracted), $M^{-1}_{ij}$ is the inverse of the migration matrix and $\Delta \rhos^{i}$ is the width of the $i$-th bin.

Once data were properly corrected,  the value of \mtPole was determined by fitting the unfolded \n3-distribution with the NLO+PS prediction from \Ref{Alioli:2011as} using the least-squares method. Table \ref{tab:n3_theory} shows the predicted values of \n3 (\Rtheory) in bins of \rhos for different \mtPole values. The fit minimized a $\chi^{2}$ defined as:

\begin{equation}
\chi^{2}=\sum \limits_{ij} (\Rcor_{i}-\Rtheory_{i}(\mtPole))V^{-1}_{ij}(\Rcor_{j}-\Rtheory_{j}(\mtPole)),
\label{eq:chi}
\end{equation}
where $\Rcor_i$ is the data value in the $i$-th bin of the corrected \n3-distribution and  $V^{-1}$ is the inverse of the statistical covariance matrix of the unfolded \n3-distribution. The quantity $\Rtheory_{i}(\mtPole)$ represents the theoretical prediction for the $i$-th bin and contains the dependence on the top-quark pole mass. The covariance matrix is obtained by producing a sample of 500 pseudo-experiments scattered around the measured values of the \n3-distributions assuming Gaussian statistical errors. Each of these pseudo-\n3-distributions was then corrected following the unfolding procedure described above and finally the covariance matrix ($V$)  was evaluated accordingly. 

\begin{table}[t]
\begin{center}
\begin{tabular}{l|l|l|l|l|l} 
 & \multicolumn{5}{c}{\Rtheory}\\
\hline
\backslashbox[25mm]{\rhos($m_0=170~\GeV$)}{\mtPole} & $170~\GeV$ & $172.5~\GeV$ &  $175~\GeV$ & $177.5~\GeV$ & $180~\GeV$ \\
\hline                                                            
0     to 0.25   & 0.1327(9) & 0.1390(44) & 0.1425(8) & 0.1487(10) & 0.1548(6)  \\
0.25  to 0.325  & 1.104(14)  & 1.134(6) & 1.172(13) & 1.208(15) & 1.251(9) \\
0.325  to 0.425 & 1.972(9)  & 2.027(4) & 2.070(9) & 2.130(11) & 2.185(7) \\
0.425  to 0.525 & 2.506(12)  & 2.561(6) & 2.587(11) & 2.644(13) & 2.674(8) \\
0.525 to 0.675  & 2.143(8)  & 2.125(4) & 2.116(7) & 2.085(9) & 2.060(6)  \\
0.675 to 1.0    & 0.353(2)  & 0.316(1) & 0.287(2) & 0.252(2) & 0.223(1) \\
\end{tabular}
\end{center}
\caption{  The \n3-distribution calculated using generated \ttbaronejet samples at NLO+PS accuracy for different \mtPole values at parton level (corresponding to \Rtheory in eq. \ref{eq:chi}). The quoted uncertainties in parentheses reflect the statistical precision of the calculation.
  \label{tab:n3_theory}}
\end{table}

The inferred \mtPole value is the one which minimizes the $\chi^{2}$ in \Eq{eq:chi} calculated by considering all bins except for the least sensitive one ($0 \leq \rhos < 0.25$). This is done because the \n3-distribution is constrained by the normalization condition. The extracted mass value does not significantly depend on this bin choice. The selected configuration is the one which gives the highest expected precision. The statistical uncertainty is taken as the mass shift that increases the $\chi^{2}$ by one unit with respect to the minimum ($\Delta \chi^2 =$ +1). The possible impact of non-perturbative effects near the threshold for \ttbaronejet production ($\rhos\sim 1$) was also studied by choosing a restricted range of $\rhos$ from 0 to 0.9 and by studying the results for each bin independently. No significant effect is observed. 

\begin{figure}[h]
\begin{center}
\includegraphics[width=0.8\columnwidth]{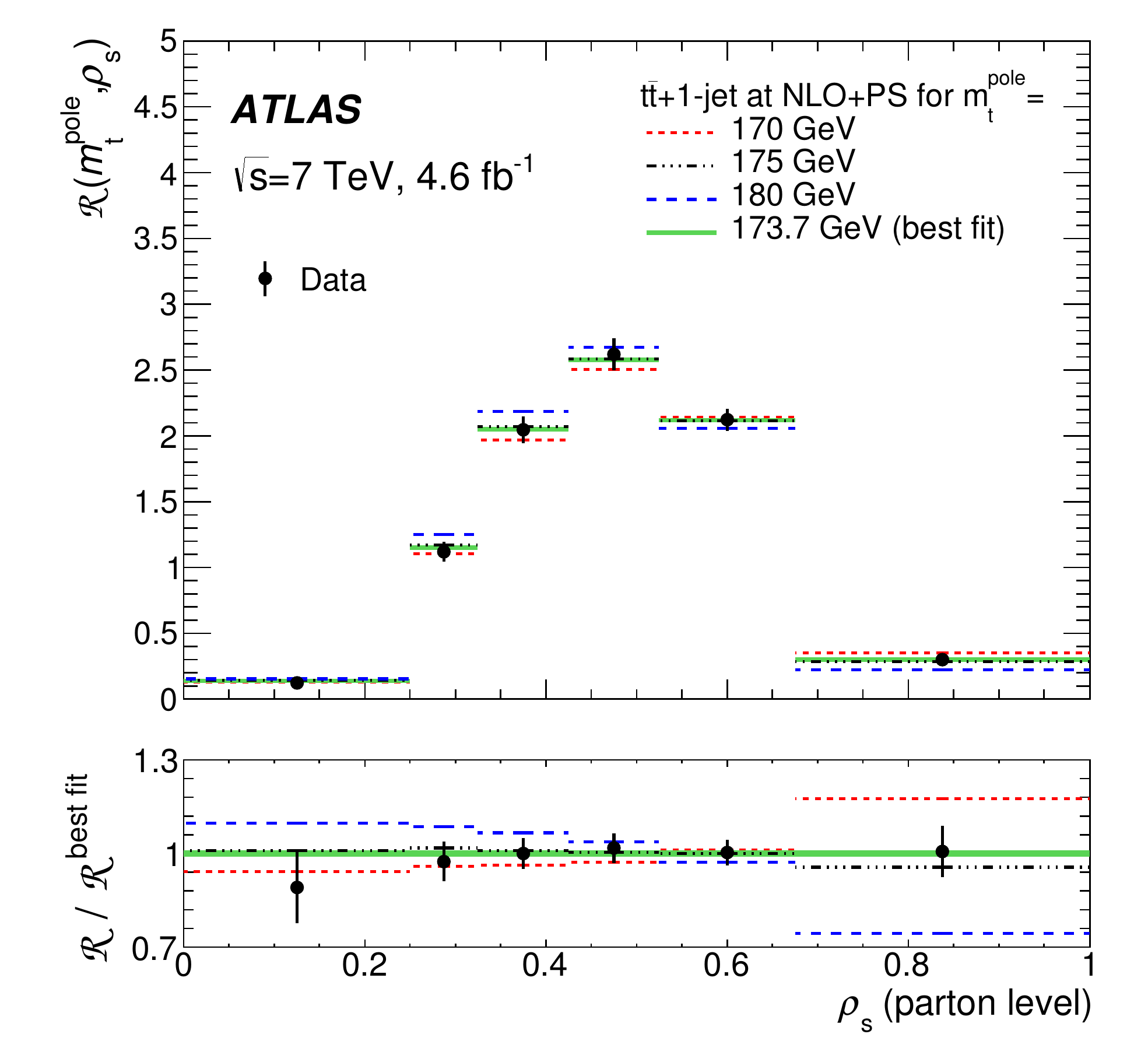}
\end{center}
\caption{\n3-distribution at parton level corrected for detector and hadronization effects after the background subtraction as a function of \rhos($m_0=170~\GeV$). The predictions of the \ttbaronejet calculation at NLO+PS using three different masses (\mtPole=170, 175 and 180~\GeV) are shown together with the result of the best fit to the data, \mtPole=173.7$\pm 1.5$ (stat.)~\GeV. The black points correspond to the data. In the lower part of the figure, the ratios of the different \n3-distributions to the one corresponding to the best fit are shown. The shaded area indicates the statistical uncertainty.
\label{fig:best_fit}}
\end{figure}

The corrected \n3-distribution is shown in figure~\ref{fig:best_fit}. For comparison purposes the predictions for three top-quark pole mass values are also shown (\mtPole=170, 175 and 180~\GeV) together with the top-quark mass extracted from the best fit, which is  \mtPole=173.7$\pm 1.5$ (stat.)~\GeV. Only the statistical uncertainty is shown in this figure.

The corrections for detector and hadronization effects based on the MC simulation might introduce a dependence on the input top-quark mass assumed in the generator. The possible impact was quantified by generating input \ttbaronejet distributions using fully simulated samples with \powheg{} + \pythia{} and with different MC masses ranging from 167.5~\gev{} to 180~\gev{}, i.e. keeping the same nominal MC parameter set except for the input top-quark mass. Each of the corresponding distributions is unfolded with the same procedure as used for data, fixing the migration matrix and correction factors, which are defined for a fixed top-quark mass of \mt= 172.5 \gev{}. Fits to the resulting \n3-distributions are performed using the parameterised mass dependence of the theoretical predictions obtained from the MC samples with input top-quark masses between 167.5 GeV and 180 GeV. Each top-quark mass extracted is compared with the corresponding input top-quark mass. Figure~\ref{fig:mcmass} shows the difference between the input and fitted masses as a function of the input mass. In the range studied here, all fit results are compatible with the input values within their statistical uncertainties.

\begin{figure}[h]
\begin{center}
\includegraphics[width=0.8\columnwidth]{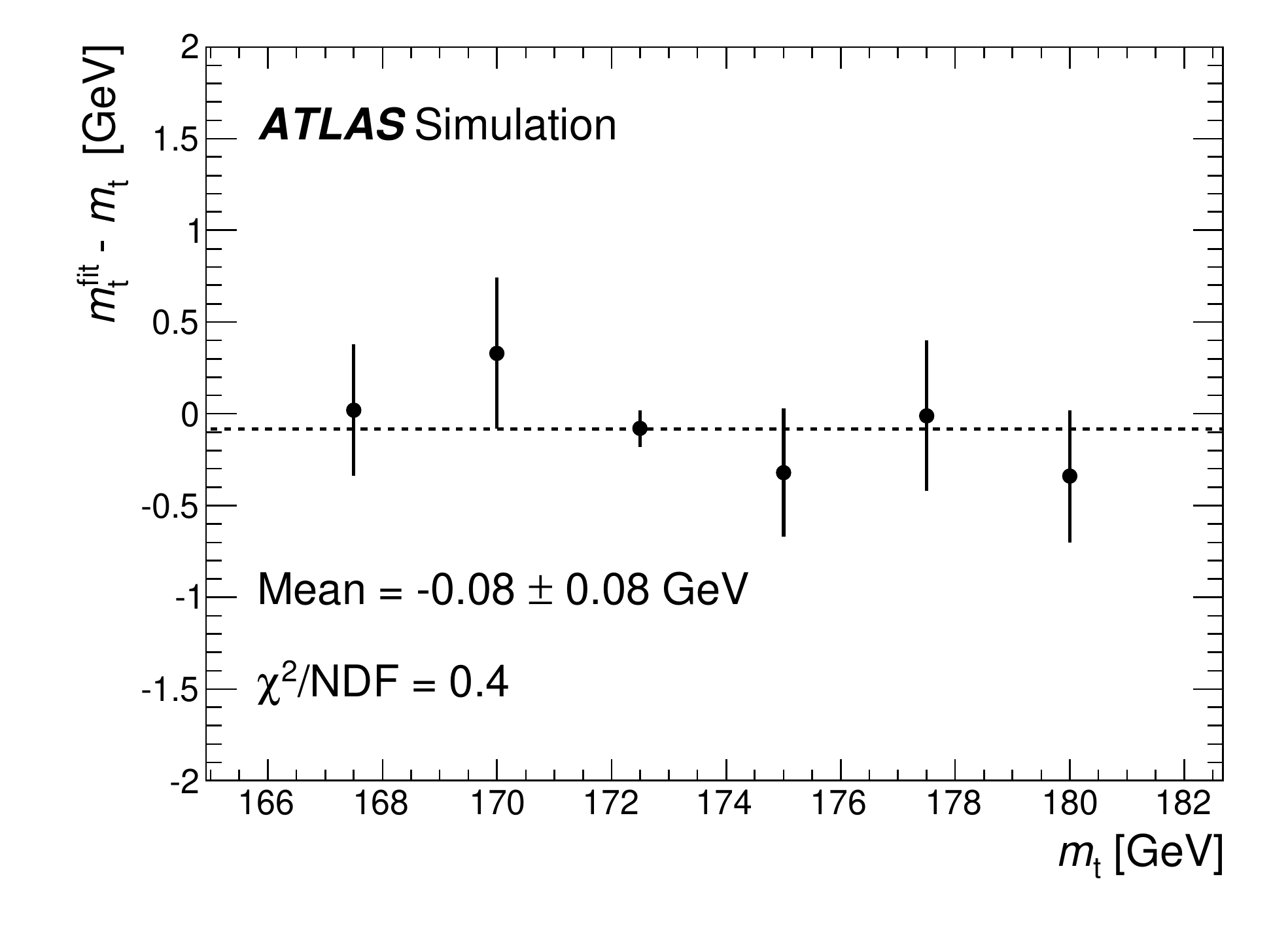}
\end{center}
\caption{Difference between the fitted mass and the top-quark mass assumed in the generated \ttbar MC predictions (\powheg+\pythia) including full detector simulation as a function of the input mass. The same unfolding procedure employed for data is performed in this study. The migration matrix and correction factors are defined for a fixed top-quark mass of \mt= 172.5 \gev{}. 
The fit is performed using the  parameterised mass dependence of the theoretical predictions obtained from the MC samples. A fit to a straight line including the point at 172.5 GeV is performed. The obtained mean value and $\chi^2$/NDF are shown. 
\label{fig:mcmass}}
\end{figure}

Existing generated samples with \powheg{} + \herwig{} including full ATLAS simulation were used to make the correction of the data without using the intermediate state of the $\ttbar$+g system. This cross-check allowed investigations of potential biases introduced by this  step. When using this sample  the correction procedure was tested including and excluding the $\ttbar$+g intermediate state. The two methods gave compatible results within $\sim$0.1 GeV for \mtPole, well within the statistical precision of the test, $\sim$0.25 GeV.

\section{Statistical and systematic uncertainties}
\label{sec:systematics}

This section describes the uncertainties that affect the extraction of the top-quark pole mass. The statistical uncertainty of the corrected result is evaluated using eq. \ref{eq:chi} using toy MC experiments to derive the covariance matrix of the fit. The additional (small) uncertainty due to the limited number of MC events used to define the unfolding procedure is evaluated by varying the migration matrix according to their statistical uncertainties. The systematic uncertainties are split into four categories: theoretical uncertainties, signal- and detector-modelling uncertainties, and finally background uncertainties. They are described in the following subsections and summarized in table \ref{tab:systematics}.

\subsection{Theoretical uncertainties}
\label{sec:systheo}

\begin{description}

\item[Scale variations:]
The calculation of \n3 is performed by setting the renormalization scale ($\muR$) equal to the factorization scale ($\muF$).  To estimate the uncertainty due to the missing higher-order terms in the calculation, these scales are varied around the central values $\mu=\muR=\muF=\mtPole$ by a factor of two up and down. The data were fitted using the predictions with a scale $\mu$  twice or half the nominal value ($\mu = 2 \mtPole$, $\mu = \mtPole/2$). The alternative choices for the scale lead to a 0.44~\gev{} lower value for the top-quark pole mass for $\mu = 2 \mtPole$ and to a 0.93~\gev{} higher value for $\mu = \mtPole/2$.

In principle one could also vary the factorization scale independently from the renormalization scale. This exercise was considered in \Ref{Alioli:2011as} and the results obtained showed very good agreement with those from the restricted scale variation considered in \Ref{Dittmaier:2008uj}. As a consequence no significant changes in the estimation of this uncertainty are expected when considering an independent variation of the scales. 

It is often argued that for normalized cross sections the method described above to evaluate the effect of uncalculated higher order terms in the perturbative calculations might be unrealistic and reduce its dependence due to cancellations of \alphas in the ratio. Therefore a different approach to evaluate this uncertainty was considered in \Ref{Alioli:2013mxa}. It consisted of expanding the \n3-distribution in powers of \alphas, thus avoiding the ratio. It was found that the two methods gave consistent estimates of the uncertainty.

In addition to these cross-checks the size of the NLO correction with respect to the LO was also computed for \n3. The study compared the LO+PS prediction using a fixed scale $\mu=m_t$ and a variable dynamic scale $\mu=\sqrt{m_t^2 + p_{{\rm T},t}^2}$ to the NLO+PS prediction with fixed scale. Differences in the range from 0.6 to 0.8 GeV were observed. The small NLO correction indicates that the calculation converges well.    

\item[Proton PDF and $\alphas$:] The uncertainties on the proton PDF and on the value of the strong coupling constant \alphas used 
in the \ttbaronejet calculation are propagated by fitting to various \n3-distributions implemented in NLO+PS calculations using different PDF sets with different $\alphas$ values. 
The central CT10~\cite{Gao:2013xoa}, MSTW2008nlo90cl~\cite{Martin:2009iq,Martin:2009bu} and NNPDF~\cite{Ball:2012cx} PDF sets were employed to estimate this uncertainty.
For each of these sets the central value of the resulting top-quark mass was calculated and the uncertainty due to the PDF corresponds to half of the maximum difference. 
The impact of varying only $\alphas$ was estimated and found to be very small, $\sigma\left( \mtPole\right)=$ 0.01~\GeV~ (for $\Delta \alphas=\pm0.002$ and the CT10 PDF set) 
since the dependence of \n3 on $\alphas$  nearly vanishes due to the use of a normalized differential distribution.

\end{description}

The total theoretical uncertainty is the sum in quadrature of the contributions
of scale variations and the 
PDF and \alphas uncertainties.

\subsection{Detector modelling}
\label{sec:sysexp}

The uncertainties on the reconstruction 
efficiency and the energy measurement of basic 
reconstructed objects (leptons, \met and jets) are propagated
to the uncertainty on the value of the top-quark mass. Variations of
all these quantities by $\pm1$ standard deviation are
implemented in MC samples that are then unfolded using the nominal response matrix. A fit to the resulting \n3-distribution is performed and the top-quark mass is extracted.
In the following --- unless otherwise stated --- the systematic uncertainties arising from the different modelling sources are calculated as half of the difference between the upward and downward variations.

\begin{description}

\item[Jet energy scale (JES) and $b$-jet energy scale:]
To estimate the impact of the jet energy scale uncertainty on the
result, the jet energy is scaled up and down by its 
uncertainty for 21 uncorrelated components which are considered 
separately~\cite{Aad:2014bia,Aad:2011he}. 

These are the experimental sources of uncertainty
with the largest impact on the precision of the mass determination.

\item[Jet energy resolution and jet reconstruction efficiency:] The effect of the jet energy resolution uncertainty is evaluated by 
smearing, before the event selection, the energy of the jets by a 
Gaussian function with a width chosen in agreement with the jet energy
resolution uncertainty.  The effect of the jet reconstruction efficiency uncertainty is evaluated by randomly discarding a fraction of jets 
from the events before the selection~\cite{jer2011}.

\item[$b$-tagging efficiency and mistag rate:] Differences in the $b$-tagging efficiency as well as $c$-jet and light-jet mistag rates 
in data and simulation are parameterized using correction factors, which are functions of \pt and \eta. These corrections are derived from data including \ttbar events
and they are varied by their uncertainties (see ~\Refs{ATLAS-CONF-2012-040, ATLAS-CONF-2012-043, ATLAS-CONF-2012-097}).
Similarly to the JES uncertainty, the uncertainty on the correction factors for the $b$-tagging 
efficiency is separated into several uncorrelated components. The systematic uncertainty is assessed
by changing the correction factor central values by $\pm1\sigma$ for each component, and performing the mass extraction. The final uncertainty due to the $b$-tagging efficiency is calculated as the quadratic sum of all contributions. 

\item[Lepton identification and lepton energy resolution:] The correction 
factors applied to the lepton identification are measured 
by comparing high-purity events from simulation and data including $Z$, $W$ and $J/\psi$ decays for 
electrons~\cite{Aad:2014fxa}, and  $Z$, $W$, $J/\psi$ and $\Upsilon$ decays for muons~\cite{Aad:2014zyy}. For the measurement of the lepton energy 
or momentum scale uncertainties, a similar procedure is used. The systematic uncertainties associated to each correction factor are estimated by changing its central value by  $\pm1\sigma$ and 
applying the mass extraction procedure. The quadratic sum of all contributions gives the quoted final uncertainty due to this source.  

\item[Modelling of the \met:] Uncertainties on the energy scale of jets or leptons are also propagated to the uncertainty of the \met. Other contributions to this uncertainty originate from 
the energy scale and resolution of the soft calorimeter energy deposits which are not included in the reconstructed jets and leptons, and contribute only to the estimation of \met. The \met
is scaled up and down by its uncertainty. The effect of these changes is propagated to the simulated events allowing to evaluate the impact on the top-quark mass measurement following the same 
procedure as for the rest of systematic uncertainties.

\end{description}

\subsection{Signal modelling}
\label{sec:sysmod}

The signal modelling uncertainties originate from: the choice of matrix element, the parton shower and hadronization model, and the choice of the PDF set used in the simulation of  
\ttbar events. In addition, uncertainties on the modelling  
of the initial- and final-state QCD radiation (ISR/FSR),  of colour reconnection, and of the 
underlying event are also accounted for.
Their impact on the extracted mass is estimated
using alternative MC samples. The alternative \n3-distribution samples are 
corrected using the nominal response matrix and the deviation from the
result of the nominal MC sample is used to estimate the uncertainty.

\begin{description}

\item[MC generator and hadronization:]  The uncertainty associated 
with the choice of MC generator is evaluated by comparing two NLO MC generators 
interfaced to the same parton shower and hadronization program: \powhegbox{} \cite{Frixione:2007vw,Alioli:2010xd} and \mcatnlo{}~\cite{Frixione:2010fsp}
both interfaced to \herwig{} are compared. The difference between the extracted masses is
taken as the generator uncertainty.

The uncertainty associated with the hadronization is 
estimated by comparing the results obtained 
with \powheg{} interfaced to either \pythia{} or \herwig{}. The full difference
is quoted as the hadronization uncertainty.

\item[Initial- and final-state radiation (ISR/FSR):] The effect of the 
ISR and FSR modelling uncertainties is evaluated by comparing 
two simulated signal samples with varied radiation 
settings. The samples to evaluate the ISR/FSR uncertainty are generated 
with \alpgen (v2.13) \cite{alpgen}+\pythia{}, which is a multileg MC generator that generates, at LO,
\ttbar plus up to five partons. The samples to estimate the ISR/FSR uncertainty correspond to variations of 
the {\em KTFAC} parameter in \alpgen{} between a factor two up and down of its nominal value (with the Perugia 2011 radLow and radHi tunes respectively~\cite{Skands:2010ak}). This
parameter determines the scale at which \alphas is evaluated for additional gluon emissions and the size of the variation considered is compatible with the measurements of additional jet activity in \ttbar events~\cite{ATLAS:2012al}. The ISR/FSR uncertainty is evaluated by taking half the 
difference between the fitted top-quark masses from the two samples. 

\item[Colour reconnection and underlying event:] The impact of 
the uncertainties in the MC models describing colour reconnection 
and the underlying event is estimated by comparing several \powheg{} MC 
samples with different tunes.
The effect of the colour reconnection
modelling uncertainty is estimated as the difference between the result 
obtained with the nominal
\powheg{} sample with the Perugia 2012 (P2012) tune and an alternative sample 
with the Perugia 2012 loCR tune~\cite{Skands:2010ak}. 
To estimate the uncertainty on
the underlying event modelling, the Perugia 2012 mpiHi tune~\cite{Skands:2014spz} is compared 
with the P2012 tune. In both cases the full mass difference from the default value is taken as the 
systematic uncertainty.

\item[Proton PDF:] Uncertainties on the proton PDF give rise to uncertainties on the efficiency of the basic event selection. These uncertainties are calculated following the 
PDF4LHC recommendations~\cite{Botje:2011sn} using \ttbar events simulated by \mcatnlo{} interfaced to \herwig{}. This uncertainty accounts for the effects of the PDF on the theoretical modelling of the \ttbar system, the hadronization and the experimental data analysis. To a large extent the first contribution is already considered in the evaluation of the theoretical uncertainties of section \ref{sec:systheo}.  In the present work the two uncertainties are considered independently. This evaluation of the total PDF uncertainty is therefore regarded as a conservative approach.  Using the values from table \ref{tab:systematics} and considering both uncertainties either completely correlated or uncorrelated changes the overall uncertainty on the PDF from 0.54 GeV to 0.58 GeV which has a rather minimal impact.

\end{description}

\subsection{Background modelling}
\label{sec:sysbkg}

The uncertainty on the background yield is taken into account by varying the 
normalization and the shape of the distributions of several contributing processes. For both $W$+jets and $Z$+jets production, 
the uncertainty on the normalization is studied following the recommendations in \Ref{wjetsnorm}. For the 5-jet final state, a total uncertainty of $54\%$ is assessed~\cite{wjetsnorm}.
For the $W$+jets background, the shape uncertainties due to the events with jets originating from heavy-flavour quarks are studied by varying the fraction of 
these events in the sample. The evaluation for this analysis follows the method described in \Ref{ATLAS:2012an}. The shape and normalization uncertainties on the misidentified and non-prompt lepton component are propagated to the top-quark mass. The most important background topologies originate from single-top plus jets production. The impact on the top-quark mass is estimated by comparing the
nominal yield (obtained using the \powheg{} generator interfaced to \pythia{}) with the equivalent result with a different set of generators (\mcatnlo{} simulation for the $s$- and $Wt$-channels 
and \acer{}~\cite{Kersevan:2004yg} for the generation of $t$-channel events). The effect of the (MC) top-quark mass used in the single-top 
background evaluation is also estimated by using two different input masses: $172.5$ and $175$~$\gev$, as well as differences in the kinematics of the single top events. The uncertainty quoted as {\it background} in table \ref{tab:systematics} is calculated as the quadratic sum of all the above contributions. 

Table \ref{tab:systematics} summarizes all uncertainties on the estimated top-quark pole mass.

\begin{table}[h]
\begin{center}
\begin{tabular}{||l|l||l|}
\hline
Description & \hfill Value & \hfill \% \\
 & \hfill $[\GeV]$ & \\
\hline
{\large\bf \mtPole } & \hfill {\large\bf 173.71}  & \\
{\bf Statistical uncertainty }              & \hfill {\bf 1.50} &  \hfill {\bf 0.9} \\
\hline
\hline
Scale variations                    & \hfill   (+0.93, $-$0.44) &  \hfill   (+0.5, $-$0.3) \\
Proton PDF (theory) and \alphas                    & \hfill  0.21  &  \hfill  0.1  \\
\bf {Total theory systematic uncertainty}                    & \hfill  \bf {(+0.95, $-$0.49)} &  \hfill \bf{(+0.5, $-$0.3)} \\
\hline
\hline
Jet energy scale (including $b$-jet energy scale)      & \hfill 0.94  &  \hfill 0.5   \\
Jet energy resolution                             & \hfill 0.02    &  \hfill  < 0.1 \\
Jet reconstruction efficiency                             & \hfill 0.05  &  \hfill < 0.1   \\
$b$-tagging efficiency and mistag rate                           & \hfill 0.17  &  \hfill 0.1   \\
Lepton uncertainties               & \hfill 0.07  &  \hfill < 0.1   \\
Missing transverse momentum           & \hfill 0.02  &  \hfill 0.1   \\
\hline
MC statistics                 & \hfill 0.13  &  \hfill < 0.1 \\
Signal MC generator                       & \hfill 0.28    &  \hfill 0.2  \\
Hadronization                             & \hfill 0.33 &  \hfill 0.2  \\
ISR/FSR                            & \hfill 0.72   &  \hfill 0.4   \\
Colour reconnection                              & \hfill 0.14  &  \hfill < 0.1 \\
Underlying event                              & \hfill 0.25 &  \hfill 0.1  \\
Proton PDF (experimental)                            & \hfill 0.54 &  \hfill 0.3 \\
\hline
Background                      & \hfill 0.20  &  \hfill 0.1   \\
\bf {Total experimental systematic uncertainty  }    & \hfill  {\bf 1.44 } &  \hfill {\bf 0.8} \\
\hline
\hline
\bf\large {Total uncertainty}                    & \hfill \bf\large{(+2.29, $-$2.14)} &  \hfill \bf\large{(+1.3, $-$1.2)} \\
\hline
\end{tabular}
\end{center}
\caption{Value of the inferred top-quark pole mass and breakdown of their associated uncertainties. \label{tab:systematics}}
\end{table}

\subsection{Studies on the definition of the extra jet}
\label{sec:sysExJet}

The extra jet of the \ttbaronejet system is required to have a \pt larger than 50 GeV but other possibilities were also investigated. The full analysis was repeated with the  \pt of the extra jet satisfying different conditions such as  \pt > 30 \gev{} and \pt > 40 \GeV. The results differ from the baseline central value by less than 0.1 GeV but the change in the systematic and statistical precision of the measurements was significant. As  the \pt requirement was decreased some systematics uncertainties, such as that due to the JES, increased and the statistical uncertainty became smaller. The original \pt condition of 50 GeV represents a good compromise for the overall balance of these uncertainties and therefore was used as the baseline of the analysis.

\subsection{Uncertainties on the measured \n3-distribution}
\label{sec:sysR}

The experimental uncertainties of the unfolded \n3-distribution for each interval of \rhos are listed in table~\ref{tab:systematicsR}. These uncertainties are obtained following the same methodology as described in previous sections for the top-quark mass but this time applied to the \n3-distribution. For each bin of the \n3-distribution, all uncertainties have been added in quadrature.

\begin{table}[h]
\begin{center}
\begin{tabular}{|c|c|c|c|}
\hline                                     
  \rhos interval & \n3 & Stat. Unc. (\%) & Syst. Unc. (\%)\\
\hline                                     
0     to 0.25         & \hfill  0.126  &  12.8  &    7.1  \\
0.25  to 0.325    & \hfill  1.122  &    6.6 &     4.5  \\
0.325  to 0.425 & \hfill  2.049  &     5.0  &    3.5  \\
0.425  to 0.525 & \hfill  2.622  &     4.6  &    2.1  \\
0.525 to 0.675  & \hfill  2.125  &     4.1  &    3.1  \\
0.675 to 1.0       & \hfill  0.302  &     8.2  &    8.1  \\
\hline                                                                       
\end{tabular}
\end{center}
\caption{Measured values of the \n3-distribution and their experimental uncertainties in percent. The statistical uncertainties are derived from the covariance matrix of eq. \ref{eq:chi}.
\label{tab:systematicsR}
}
\end{table}

\section{Results and discussion}
\label{sec:results}

This paper describes an experimental measurement of the top-quark mass using the novel method proposed in \Ref{Alioli:2013mxa}. The value of \mtPole is obtained from a fit to the normalized differential cross section $\n3(\mtPole,\rhos)$ for \ttbar production with at least one extra jet, \ttbaronejet, as a function of the inverse of the invariant mass of the \ttbaronejet system, \rhos. This method allows a rigorous theoretical interpretation of the extracted mass parameter in terms of the top-quark pole mass or the running mass in the \MSbar scheme. In the present analysis only the top-quark pole mass (\mtPole) is measured, although future studies should also be able to determine the running mass when the theoretical calculations become available. 

Events with the $\ttbaronejet$ final state were selected using 4.6 fb$^{-1}$ of 7 TeV $pp$ collision data collected by the ATLAS experiment at the LHC in 2011. The total background in the \ttbaronejet  sample is $\sim$6\%.  Many distributions were studied to demonstrate the overall agreement between the MC predictions and data. A thorough study of the systematic effects with impact on the measurement was carried out and the associated uncertainties quantified. Experimental systematic uncertainties were computed for detector, signal and background modelling.

The measured top-quark pole mass is:

\begin{flalign*}
  \mtPole&=173.7 \pm 1.5 ~({\rm stat.}) \pm 1.4 ~({\rm syst.}) ^{+ 1.0}_{-0.5} ~({\rm theory})~{\rm GeV},
\end{flalign*}
where the theoretical uncertainties include the uncertainty due to missing higher orders in 
the perturbative NLO calculation, as well as uncertainties due to the PDF and $\alphas$ used in the calculations. The experimental uncertainty accounts for the 
uncertainties due to the imperfections in the modelling of the detector response, the background yield and the uncertainties arising from the 
signal modelling including hadronization. The dominant experimental uncertainties are due to the jet energy calibration
($0.94~\GeV$) and the initial- and final- state radiation modelling ($0.72~\GeV$).

This measurement constitutes the first extraction of the top-quark pole mass 
from a measurement of the differential \ttbaronejet production 
cross section as a function of the inverse of the invariant mass of the \ttbaronejet
system. It represents the most precise measurement 
of the top-quark pole mass to date with a total uncertainty of +2.3 GeV and -2.1 GeV. The value obtained for the top-quark pole mass agrees 
with the most accurate previous top-quark mass measurement in the pole-mass scheme~\cite{Aad:2014kva} and with the direct top-quark 
mass measurement~\cite{ATLAS-CONF-2014-008}.

The analysis presented in this paper is statistically uncorrelated from the \mtPole measurement using the inclusive cross-section measured in dilepton events \cite{Aad:2014kva}. The measurements could therefore potentially be combined, however this would require a detailed study of the correlations of both the uncertainties on the experimental measurements and the theoretical calculations.


\section*{Acknowledgements}

We are very grateful to S.~Alioli, S.Moch and P.~Uwer for their fruitful collaboration in developing the theoretical aspects of the analysis presented in this paper and for their continuous support in implementing the new methodology in the experimental analysis.

We thank CERN for the very successful operation of the LHC, as well as the
support staff from our institutions without whom ATLAS could not be
operated efficiently.

We acknowledge the support of ANPCyT, Argentina; YerPhI, Armenia; ARC,
Australia; BMWF and FWF, Austria; ANAS, Azerbaijan; SSTC, Belarus; CNPq and FAPESP,
Brazil; NSERC, NRC and CFI, Canada; CERN; CONICYT, Chile; CAS, MOST and NSFC,
China; COLCIENCIAS, Colombia; MSMT CR, MPO CR and VSC CR, Czech Republic;
DNRF, DNSRC and Lundbeck Foundation, Denmark; EPLANET, ERC and NSRF, European Union;
IN2P3-CNRS, CEA-DSM/IRFU, France; GNSF, Georgia; BMBF, DFG, HGF, MPG and AvH
Foundation, Germany; GSRT and NSRF, Greece; ISF, MINERVA, GIF, I-CORE and Benoziyo Center,
Israel; INFN, Italy; MEXT and JSPS, Japan; CNRST, Morocco; FOM and NWO,
Netherlands; BRF and RCN, Norway; MNiSW and NCN, Poland; GRICES and FCT, Portugal; MNE/IFA, Romania; MES of Russia and ROSATOM, Russian Federation; JINR; MSTD,
Serbia; MSSR, Slovakia; ARRS and MIZ\v{S}, Slovenia; DST/NRF, South Africa;
MINECO, Spain; SRC and Wallenberg Foundation, Sweden; SER, SNSF and Cantons of
Bern and Geneva, Switzerland; NSC, Taiwan; TAEK, Turkey; STFC, the Royal
Society and Leverhulme Trust, United Kingdom; DOE and NSF, United States of
America.

The crucial computing support from all WLCG partners is acknowledged
gratefully, in particular from CERN and the ATLAS Tier-1 facilities at
TRIUMF (Canada), NDGF (Denmark, Norway, Sweden), CC-IN2P3 (France),
KIT/GridKA (Germany), INFN-CNAF (Italy), NL-T1 (Netherlands), PIC (Spain),
ASGC (Taiwan), RAL (UK) and BNL (USA) and in the Tier-2 facilities
worldwide.

\bibliography{notebib_atlasstyle}   

\clearpage
\newpage



\begin{flushleft}
{\Large The ATLAS Collaboration}

\bigskip

G.~Aad$^{\rm 85}$,
B.~Abbott$^{\rm 113}$,
J.~Abdallah$^{\rm 152}$,
S.~Abdel~Khalek$^{\rm 117}$,
O.~Abdinov$^{\rm 11}$,
R.~Aben$^{\rm 107}$,
B.~Abi$^{\rm 114}$,
M.~Abolins$^{\rm 90}$,
O.S.~AbouZeid$^{\rm 159}$,
H.~Abramowicz$^{\rm 154}$
% ...
\bigskip
\\
% ...
$^{11}$ Institute of Physics, Azerbaijan Academy of Sciences, Baku, Azerbaijan\\
% ...
$^{85}$ CPPM, Aix-Marseille Universit{\'e} and CNRS/IN2P3, Marseille, France\\
% ...
$^{90}$ Department of Physics and Astronomy, Michigan State University, East Lansing MI, United States of America\\
% ...
$^{107}$ Nikhef National Institute for Subatomic Physics and University of Amsterdam, Amsterdam, Netherlands\\
% ...
$^{113}$ Homer L. Dodge Department of Physics and Astronomy, University of Oklahoma, Norman OK, United States of America\\
$^{114}$ Department of Physics, Oklahoma State University, Stillwater OK, United States of America\\
% ...
$^{117}$ LAL, Universit{\'e} Paris-Sud and CNRS/IN2P3, Orsay, France\\
% ...
$^{152}$ Institute of Physics, Academia Sinica, Taipei, Taiwan\\
% ...
$^{154}$ Raymond and Beverly Sackler School of Physics and Astronomy, Tel Aviv University, Tel Aviv, Israel\\
% ...
$^{159}$ Department of Physics, University of Toronto, Toronto ON, Canada\\
% ...
\end{flushleft}


\clearpage
\newpage


\end{document}